\documentclass[print]{aa} 

\usepackage{graphicx}
\usepackage{subfigure}
\usepackage{float}
\usepackage{amsmath}    
\usepackage{amssymb}     
\usepackage[normalem]{ulem}
\usepackage{color}
\usepackage{threeparttable}
\usepackage{hyperref}
\usepackage{txfonts}
\usepackage{newtxtext}
\begin{document} 

   \title{The propagation-induced circular polarization of fast radio bursts in relativistic plasma}
   
   \titlerunning{Propagation-induced CP of FRBs}
   \authorrunning{Zhao et al.}

   \author{Z. Y. Zhao\inst{1}
        \and
            F. Y. Wang\inst{1, 2}\thanks{E-mail: fayinwang@nju.edu.cn}
        }

    \institute{School of Astronomy and Space Science, Nanjing University, Nanjing 210093, China\\
        \and
            Key Laboratory of Modern Astronomy and Astrophysics (Nanjing University), Ministry of Education, Nanjing 210093, China\\
    }

   \date{Received xxx; accepted xxx}

  \abstract
   {Although the physical origin of fast radio bursts (FRBs) remains unknown, magnetars are the most likely candidates. The polarization properties of FRBs offer crucial insights into their origins and radiation mechanisms. Significant circular polarization (CP) has been observed in some FRBs. CP may result from intrinsic radiation or propagation effects, both within and outside the magnetosphere. Recent observations indicate that polarization properties of FRB 20201124A can change over short timescales (about tens of milliseconds), challenging models that attribute CP to out-of-magnetosphere emission and propagation. Additionally, some magnetospheric radiation models predict that bursts with high CP produced by off-axis emission will be systematically fainter, which contradicts the observations. We propose that CP arises from magnetospheric propagation effects caused by relativistic plasma. We identify the conditions under which high CP occurs, finding it to be rare. Moreover, our model accounts for the more commonly observed low CP and the varying handedness of CP.}


   \maketitle
   \nolinenumbers
%
\section{Introduction}\label{sec:intro}
Fast radio bursts (FRBs) are millisecond-duration 
radio transients with mysterious origin \citep{Xiao2021,Petroff2022,Zhang2023,Wu2024}. In the more than a decade since FRBs were discovered, the field has made many significant observational advances \citep{Lorimer2024}. Among them, the most meaningful one is the detection of FRB 200428 from a Galactic magnetar \citep{Bochenek2020,CHIME2020}, which indicates that at least some of the FRBs originate from active magnetars \citep{Lyubarsky2014,Beloborodov2017,Metzger2017,Wang2017,Metzger2019,Kumar2020,Lu2020,Wang2020}. The physical origin of FRB is still unknown, and even for the magnetar model, the radiation location and mechanism are controversial: some models claim that emission originates from the magnetosphere of magnetars (so-called `close-in' scenario, e.g., \citealt{Yang2018,Kumar2020,Lu2020,Zhang2022}), some models believe that the radiation comes from relativistic shock farther away from the central engine (so-called `faraway' scenario, e.g., \citealt{Lyubarsky2014,Beloborodov2017,Metzger2019}), or FRBs may cause by collisions between magnetars and asteroids or asteroid belts \citep{Geng2015,Dai2016,Wu2023}.
All of the above models can explain the millisecond duration, peak frequency and high luminosity. 

The polarization properties of FRBs can provide key clues to reveal their origins and radiation mechanisms. 
Initially, repeating FRBs with polarization measurements showed high linear polarization (LP, some up to almost 100\%) and a nearly constant polarization position angle (PA, see \citealt{Michilli2018,Day2020}). While circular polarization (CP) and swings of PA have been observed for some apparent one-off FRBs \citep{Masui2015,Prochaska2019,Cho2020,Day2020}. Diverse PAs were first detected in the repeating FRB 20180301A: PAs are non-varying during some bursts, while PAs swing across some pulse profiles \citep{Luo2020}. An S-shaped PA evolution has recently been detected in one-off FRB \citep{Mckinven2024}. In follow-up monitoring of some repeated FRBs, significant CP had been discovered, such as FRB 20121102A \citep{Feng2022}, FRB 20201124A \citep{Xu2022} and FRB 20190520B \citep{Anna-Thomas2023}. 

Similar to the radio pulsar, the generation of CP can be attributed to radiation mechanisms or propagation effects.
For magnetospheric models, such as coherent curvature radiation (CR) or inverse Compton scattering (ICS), LP and CP are expected for the on-axis and off-axis emission \citep{WWY2022,Tong2022,Qu2023}. The S-shape swing of PA is interpreted as the rotation vector model (RVM, \citealt{Radhakrishnan1969}). For models involving synchrotron maser emission at relativistic magnetized shocks, linearly polarized X-mode waves are generated via Particle-in-cell (PIC) simulations, which can explain the almost 100\% LP and the flat PA \citep{Metzger2019}. With the generation of O-mode, the varying PA can be achieved \citep{Iwamoto2024}. Some propagation models are proposed to explain the CP, including Faraday conversion, both in relativistic and non-relativistic plasma \citep{Gruzinov2019,Wang2022,LiDZ2023,Xia2023,Zhao2024,Lower2024}, and cyclotron/synchrotron absorption \citep{Qu2023}.

Recently, $\sim$90\% CP has been detected in FRB 20201124A \citep{Jiang2024}. They find that the polarization properties can change in a few tens of milliseconds, which disfavors the out-of-magnetosphere emission and propagation models. The popular magnetospheric radiation models (e.g., coherent CR or ICS) predict that CP, produced by off-axis emission, will be systematically fainter \citep{WWY2022,Tong2022,Qu2023}. However, no significant difference was found in the energy distribution between high CP and low CP samples \citep{Jiang2024}. An abrupt orthogonal jump of PA has recently been discovered in FRB 20201124A \citep{Niu2024}. Orthogonal jumps, observed in radio pulsars \citep{Manchester1975,Karastergiou2009}, 
have often been interpreted in terms of two orthogonal polarization modes (OMs). When waves propagate in birefringent plasma in the magnetosphere, the waves can be decomposed into two polarization eigenmodes \citep{Petrova2001}. However, recent work has suggested that the O-mode may be rapidly damped during the propagation of strong GHz waves \citep{Beloborodov2024}. The origin of such PA jumps is not the focus of the present work.

Although some fine-tuning of radiation models could try to make sense of the latest observations \citep{Jiang2024}, an explanation of magnetospheric propagation seems more plausible.  The magnetosphere propagation effect depends on the relativistic plasma dispersion function, which has been well-studied based on different plasma models or pair distribution functions. From the numerical simulations of the pair creation \citep{Hibschman2001,Arendt2002}, a relativistic thermal distribution is chosen in this work \citep{Melrose1999,MG99}. In this framework, only relativistic corrections to the magneto-ionic theory, both relativistic motion and intrinsically relativistic distribution, are required to explain the observed CP \citep{Melrose2004}. The polarization properties are determined by the wave modes in the polarization limiting region.

In this paper, we use the propagation-induced CP in magnetospheric relativistic plasma to explain the polarization of FRBs. This paper is organized as follows. 
The wave dispersion and the polarization description in cold plasma are shown in Section \ref{sec:cold}. The conditions for CP are given in Section \ref{sec:relativistic}: relativistic streaming plasma in Section \ref{sec:stream} and intrinsically relativistic plasma in Section \ref{sec:intrinsic}. The explanations of the polarization observations of FRB 20201124A, FRB 20220912A and FRB 20240114A are given in Section \ref{sec:obs}. Finally, our conclusion is given in Section \ref{sec:con}. In this work, unless otherwise noted, we use the expression $Q_x = Q/10^x$ in cgs units.

\begin{figure}
    \centering
    \resizebox{\hsize}{!}{\includegraphics{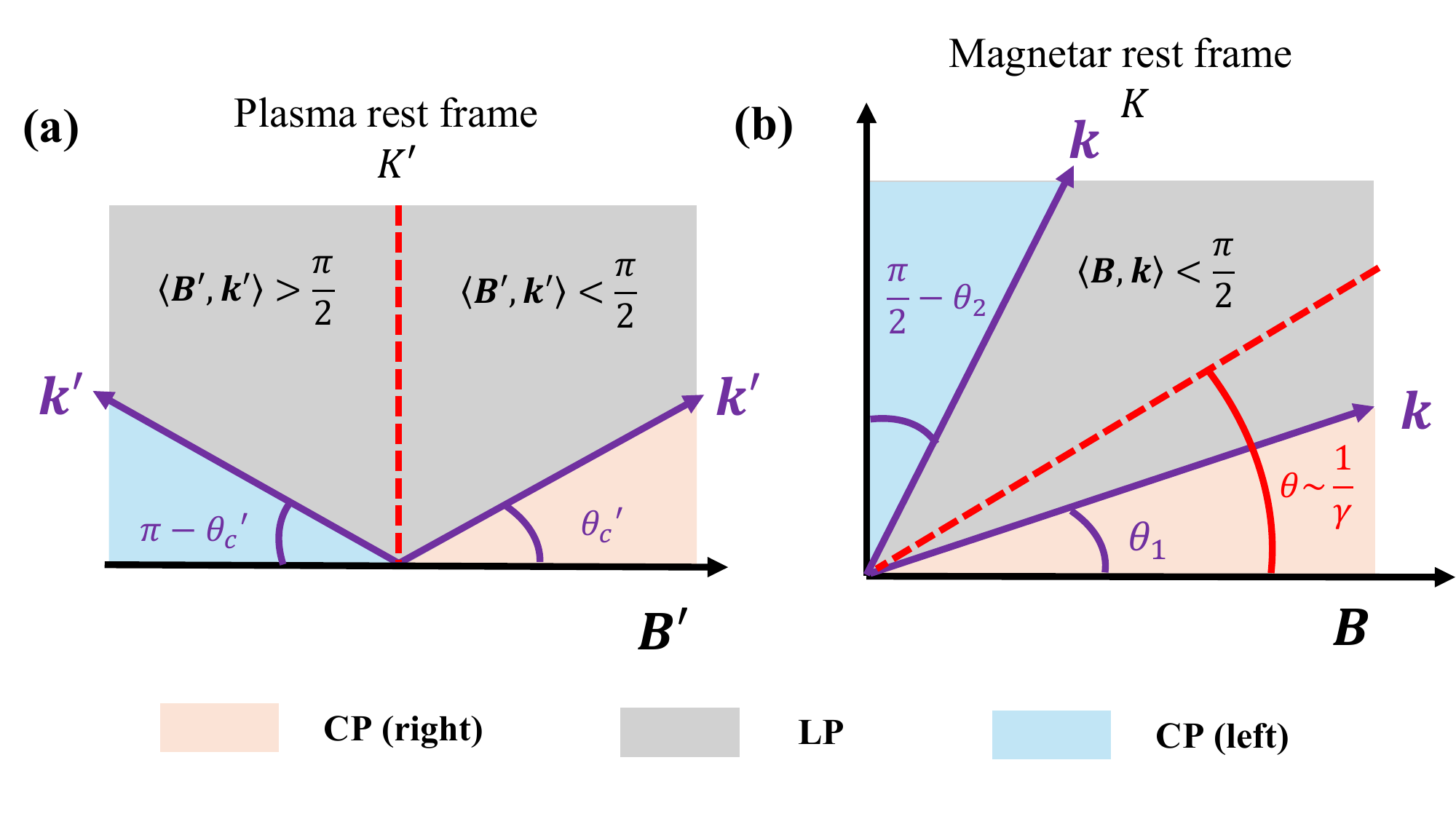}}
    \caption{The polarization degrees and the transition angles in the plasma rest frame $K^{\prime}$ (panel a) and the magnetar rest frame $K$ (panel b) for a given mode. The plasma in the magnetosphere is streaming with the bulk Lorentz factor $\gamma_{\mathrm{s}}$. The angle between the magnetic field direction and wave vector is $\theta$. Panel (a). In the plasma rest frame, for a given mode, the handedness of CP is opposite for the forward cone ($0<\theta^{\prime}<\pi/2$) and the backward cone ($\pi/2<\theta^{\prime}<\pi$). High CP occurs for $\theta^{\prime} \lesssim \theta_{\mathrm{c}}^{\prime}$, high LP occurs for $\theta_{\mathrm{c}}^{\prime} \lesssim \theta^{\prime} \lesssim \pi-\theta_{\mathrm{c}}^{\prime}$ and high CP in the opposite handedness occurs for $\pi-\theta_{\mathrm{c}}^{\prime} \lesssim \theta^{\prime} \lesssim \pi$. Panel (b). In the magnetar rest frame, the Lorentz transformation shrinks the forward cone but expands the backward cone into the forward hemisphere in the frame $K$, making it possible to have two transition angles instead of just one in non-relativistic cold plasma for forward propagation. The handedness of CP is opposite for $0<\theta<1/\gamma_{\mathrm{s}}$ and $1/\gamma_{\mathrm{s}}<\theta<\pi/2$. High CP occurs for $\theta \lesssim \theta_1$, high LP occurs for $\theta_1 \lesssim \theta \lesssim \theta_2$ and high CP in the opposite handedness occurs for $\theta_1 \lesssim \theta \lesssim \pi$/2.}
    \label{fig:trtheta}
\end{figure}

\section{The polarization in cold plasma}\label{sec:cold}
Here, we consider the propagation of waves in the cold plasma with a mixture of different particles. It is conventional to introduce the magnetoionic parameters
\begin{align}
    &X=\omega_{\mathrm{p}}^2 / \omega^2, \quad Y=\Omega_\alpha / \omega,\\
    & \eta= \frac{n_{+}-n_{-}}{n_{-}+n_{+}},
\end{align}
where $\omega$ is the wave frequency. The plasma frequency is $\omega_{\mathrm{p}}^2=4\pi e^2 n_{\mathrm{e}}/m_{\mathrm{e}}$, 
where $n_{\mathrm{e}}=n_{-}+n_{+}$ is the total electron number density, and the cyclotron frequency is $\Omega_\alpha=\left|q_\alpha\right| B_{\mathrm{bg}} / m_\alpha c$
for species $\alpha$ with $B_{\mathrm{bg}}$ being the background magnetic field strength. For electron and positron plasma, we have $\eta=0$ 
for pure pair plasma and $\eta=-1$ for pure electron gas. For ion-electron plasma, when the influence of ions is neglected, it can be that $\eta=-1$.
The polarization of a given mode can be described as transverse components $T_{ \pm}$ of the polarization vector (see Appendix \ref{app:pol}). Two orthogonal linear polarizations (LP) correspond to $T=0, \infty$ and two opposite CP correspond to $T= \pm 1$. Transverse components satisfy a quadratic equation \citep{Melrose1991}
\begin{equation}\label{eq:T}
T^2-\mathcal{R} T-1=0,
\end{equation}
where $\mathcal{R}$ is
\begin{equation}\label{eq:R_cold}
\begin{aligned}
    \mathcal{R}&=\frac{\left(P S-S^2+D^2\right) \sin ^2 \theta}{P D \cos \theta}\\
    &=\frac{Y \sin ^2 \theta}{\eta(1-X) \cos \theta}.
\end{aligned}
\end{equation}
The plasma parameters $P,S,D$ can be obtained from the dielectric tensor for cold plasma (see Appendix \ref{app:cold} for details). In equation (\ref{eq:R_cold}), we choose a background magnetic field along the 3-axis, and $\theta$ is the angle between the wave vector $\boldsymbol{k}$ and the magnetic field $\boldsymbol{B}$. The solutions of equation (\ref{eq:T}) are

\begin{equation}\label{eq:T_cold}
T=T_{ \pm}=\frac{1}{2}\left[\mathcal{R}\pm\left(\mathcal{R}^2+4\right)^{1 / 2}\right]. 
\end{equation}
The polarization ellipses of two modes satisfy $T_{+} T_{-}=-1$, which means they are orthogonal. The degree of LP is $\Pi_\mathrm{L}=\left(T_{ \pm}^2-1\right) /\left(T_{ \pm}^2+1\right)$ and the degree of CP is $\Pi_\mathrm{V}=2 T_{ \pm} /\left(T_{ \pm}^2+1\right)$. The handedness of CP is opposite when $0<\theta<\pi/2$ and $\pi/2<\theta<\pi$. The ratio of CP and LP is $|\Pi_{\mathrm{V}} / \Pi_\mathrm{L}|=2 /|\mathrm{R}|$. We can define a transition angle $\theta_{\mathrm{c}}$ for $|\Pi_{\mathrm{V}} / \Pi_\mathrm{L}|=1$, corresponding to $|\mathcal{R}|=2$. As shown in panel (a) of Figure \ref{fig:trtheta}, high CP occurs for $|\mathcal{R}|\lesssim 2$, or $\theta \lesssim \theta_{\mathrm{c}}$ and $\pi-\theta_{\mathrm{c}} \lesssim \theta \leqslant \pi$, and high LP occurs for $|\mathcal{R}|\gg 2$ or $\theta_{\mathrm{c}} \lesssim \theta \lesssim \pi-\theta_{\mathrm{c}}$. Writing $r=Y / |\eta|(1-X)$, the transition angle $\theta_{\mathrm{c}}$ is \citep{Melrose2004}

\begin{equation}  
\theta_c=\arccos \left[\frac{\left(1+r^2\right)^{1 / 2}-1}{r}\right] \approx\left\{\begin{array}{l}  
\pi / 2-r \text { for } r \ll 1, \\
(2 / r)^{1 / 2} \text { for } r \gg 1 .
\end{array}\right.  
\end{equation}

The typical ion-electron plasma in most astrophysical environments applies to high-frequency and weak magnetic field approximation ($\omega \gg \omega_{\mathrm{p}}, \omega \gg \Omega_{\mathrm{e}}$), which means $r\ll 1$ and $\theta_{\mathrm{c}}\rightarrow \pi/2$. This is the familiar result for cold plasma, the two modes are approximately circularly polarized with opposite handedness (corresponding to $T_{\mathrm{R}} \simeq+1$ and $T_{\mathrm{L}} \simeq-1$) except for $\theta\sim \pi/2$. The observed radio waves are linearly polarized due to the superposition of two orthogonal modes. Suppose the absorption of the medium is ignored, the polarization properties will only change when the wave passes through the region with the angle between the wave vector and magnetic field close to the transition angle $\theta\sim \theta_{\mathrm{c}}\sim\pi/2$ (usually the magnetic field reversal region, see \citealt{Gruzinov2019,LiDZ2023,Xia2023}). The polarization evolution of waves in cold plasma, especially for wave propagation out of the magnetosphere, has been well-studied before \citep{Gruzinov2019,Wang2022,LiDZ2023,Xia2023,Zhao2024}.

In the magnetosphere, the waves have two orthogonal modes, the ordinary mode (O-mode) and the extraordinary mode (X-mode). Due to the birefringence of magnetospheric plasma, the two modes are expected to have slightly different propagation properties \citep{Barnard1986,Petrova2000,Noutsos2015}. 
In principle, such differential propagation may contribute to changes in the observed polarization, including orthogonal PA jumps. However, recent studies have shown that the O-mode can suffer strong damping \citep{Beloborodov2024}, while \citealt{Wang2024} showed that subluminous O-mode waves can be subject to strong Landau damping. Orthogonal PA jumps are infrequent for FRBs, with \citet{Niu2024} finding only 3 cases out of 2,000 bursts from FRB 20201124A. A detailed investigation of the origin of PA jumps and mode-dependent damping is beyond the scope of this work. Here, we investigate the polarization properties of each eigenmode separately. Even if the O-mode is strongly damped during subsequent propagation, the X-mode results remain applicable, and such damping does not change our main conclusions regarding the polarization properties of the magnetospheric eigenmodes. The polarization properties are determined by the wave modes in the polarization-limiting region.

For relativistic pair plasma in the magnetosphere, the corresponding transition angle is not constant, and there may be two solutions ($\theta_1$ and $\theta_2$, see Section \ref{sec:relativistic}). 

\begin{table}
        \caption{The transition angle of relativistic plasma}
        \label{tab:trtheta}
\centering
	\begin{tabular}{ccccccc}
	\hline
        & $r$ & $\gamma_{\mathrm{s}}$ & $\rho$ &$\langle\gamma\rangle$ &$\theta_1$ & $\theta_2$ \\
         & &    &  &  & (rad) & (rad) \\
        \hline
        Relativistic & $10^6$ & 50 & & & $1.4\times10^{-6}$ & \\
        cold plasma & $10^4$ & $10^2$ &  & 
        & $5.0\times10^{-6}$ & 0.45\\
        \hline
         & $10^6$ & 50 & 1 & 1.7 &$8.6\times10^{-7}$ & \\
        Intrinsically  & $10^4$ & $10^2$ & 1 & 1.7 & $3.0\times10^{-6}$ & 0.57\\
        relativistic  & $10^4$ & $10^2$ & 0.1 & $\sim10$ & $1.1\times10^{-6}$ & 0.86\\
        thermal plasma& $10^4$ & $10^2$ & 10 & 1.05 & $4.7\times10^{-6}$ & 0.46\\
        \hline
\end{tabular}
\end{table}
\begin{figure*}
    \centering
    \resizebox{\hsize}{!}{\includegraphics{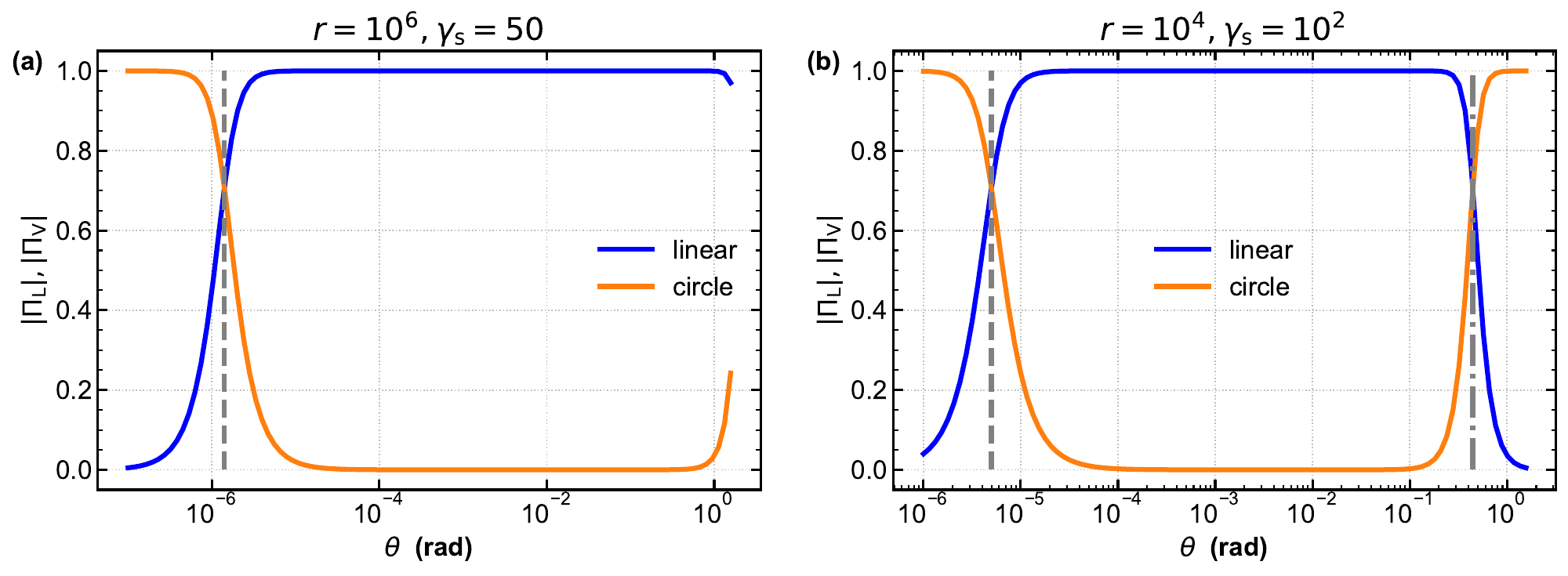}}
    \caption{The LP ($|\Pi_{\mathrm{L}}|$, blue lines) and CP ($|\Pi_{\mathrm{V}}|$, orange lines) degrees as a function of the angle between the wave vector and magnetic field. For a strong background magnetic field and low Lorentz factor (e.g., $r=10^{6},\gamma_{\mathrm{s}}=10$), the second transition angle $\theta_2$ does not exist (panel (a)). For a weak background magnetic field and large Lorentz factor (e.g., $r=10^{4},\gamma_{\mathrm{s}}=100$), there are two transition angles (panel (b) ). The transition angles $\theta_1$ and $\theta_2$ are represented by gray dashed and dotted-dashed lines, respectively. The transition angles of different parameters are listed in Table \ref{tab:trtheta}.}
    \label{fig:Pi1}
\end{figure*}

\begin{figure*}
    \centering
    \resizebox{\hsize}{!}{\includegraphics{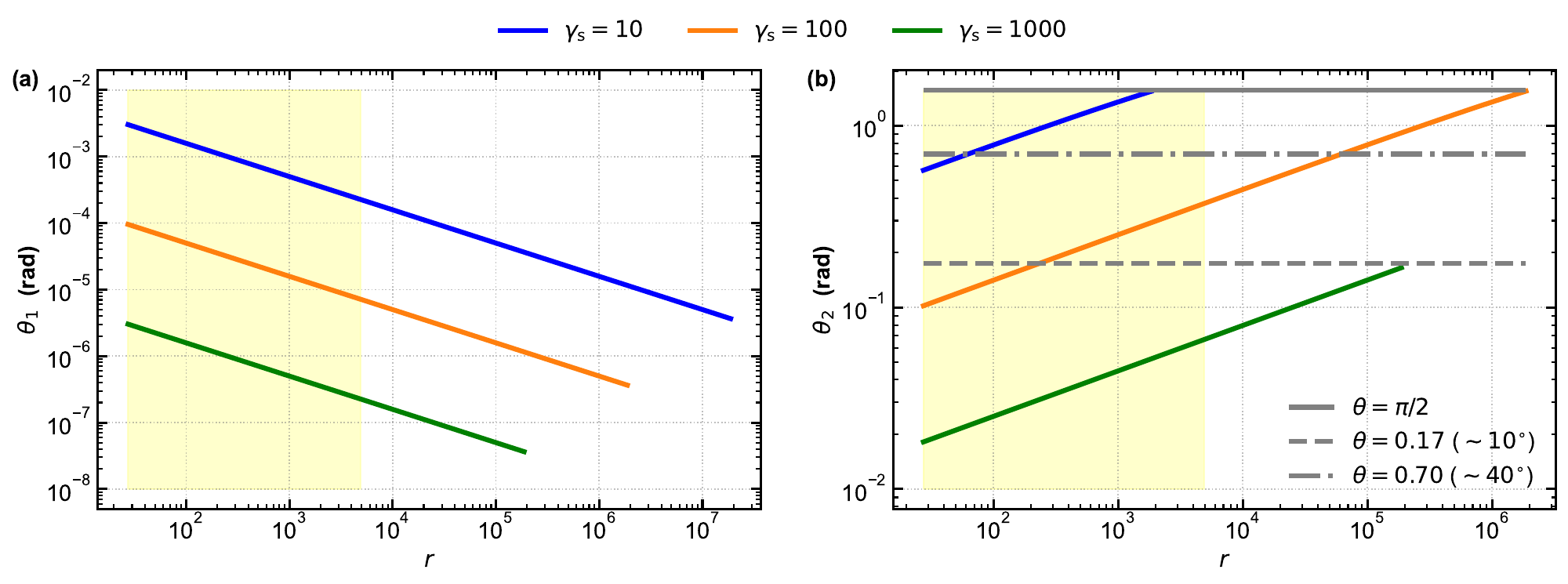}}
    \caption{The transition angles $\theta_1$ (panel (a)) and $\theta_2$ (panel (b)) for relativistic streaming cold plasma in the magnetar rest frame. If FRBs emit along the tangent of the field line, the propagation angle between the wave vector and the magnetic field can be constrained by the dipolar field geometry. The maximum of $\theta$ is $\theta_{\max}\sim10^{\circ}-40^{\circ}$ \citep{Qu2022}. The limits of the maximum angles $\theta_{\max}=90^{\circ}$, $\theta_{\max}=10^{\circ}$, and $\theta_{\max}=40^{\circ}$ are also shown in gray horizontal solid, dashed, and dotted-dashed lines. Under certain parameters, the second transition angle $\theta_2$ approaches or even exceeds the maximum angle. The shade region represents the value of $r_{\min}$ for FRBs with the isotropic luminosity of $10^{38}-10^{41}$ erg s$^{-1}$.}
    \label{fig:thetac1}
\end{figure*}

\section{The wave polarization in relativistic plasma}\label{sec:relativistic}

The wave polarization in non-relativistic ion-electron cold plasma is shown above, which is inapplicable to relativistic pair plasma in the magnetosphere. For pure pair plasma ($\eta=0$), from equations (\ref{eq:R_cold}) and (\ref{eq:T_cold}) we have $T_{\pm}=0,\infty$, which means the natural modes are completely linearly polarized. However, the net charge density usually exists in the magnetosphere \citep{Goldreich1969}. The corresponding Goldreich–Julian (GJ) density for a dipole field is
\begin{equation}\label{eq:n_GJ}  
n_{\mathrm{GJ}} \simeq \frac{\Omega B_{\mathrm{bg}}}{2 \pi e c} \simeq6.9 \times 10^7 \mathrm{~cm}^{-3} B_\mathrm{s, 15} P^{-1} \hat{r}_2^{-3},
\end{equation}
where $B_{\mathrm{s}}$ is the surface magnetic field, $\hat{r}=r_0/R_{\star}$ and $R_{\star}$ is the magnetar radius. $\Omega=2 \pi / P$ is the angular frequency with $P$ being the rotation period. The number density of pairs in the magnetosphere can be estimated as $\kappa n_{\mathrm{GJ}}$ due to electron–positron pair production, where $\kappa$ is the multiplicity. The study of an electromagnetic cascade implies that the plasma is streaming relativistically in the magnetosphere, with the bulk Lorentz factor $\gamma_{\mathrm{s}}\sim 10^2-10^3$ and the multiplicity factor $\kappa\sim 10^2-10^5$\citep{Daugherty1982,Arendt2002,Timokhin2015}. For a magnetar, the multiplicity factor is expected to be $\kappa\sim 1-10^2$ \citep{Medin2010,Beloborodov2013}. If the multiplicity is $\kappa\gg 1$, $\eta$ can be be approximate to $\eta\sim1/\kappa$. 

\begin{figure*}
    \centering
    \includegraphics[width = 1\textwidth]{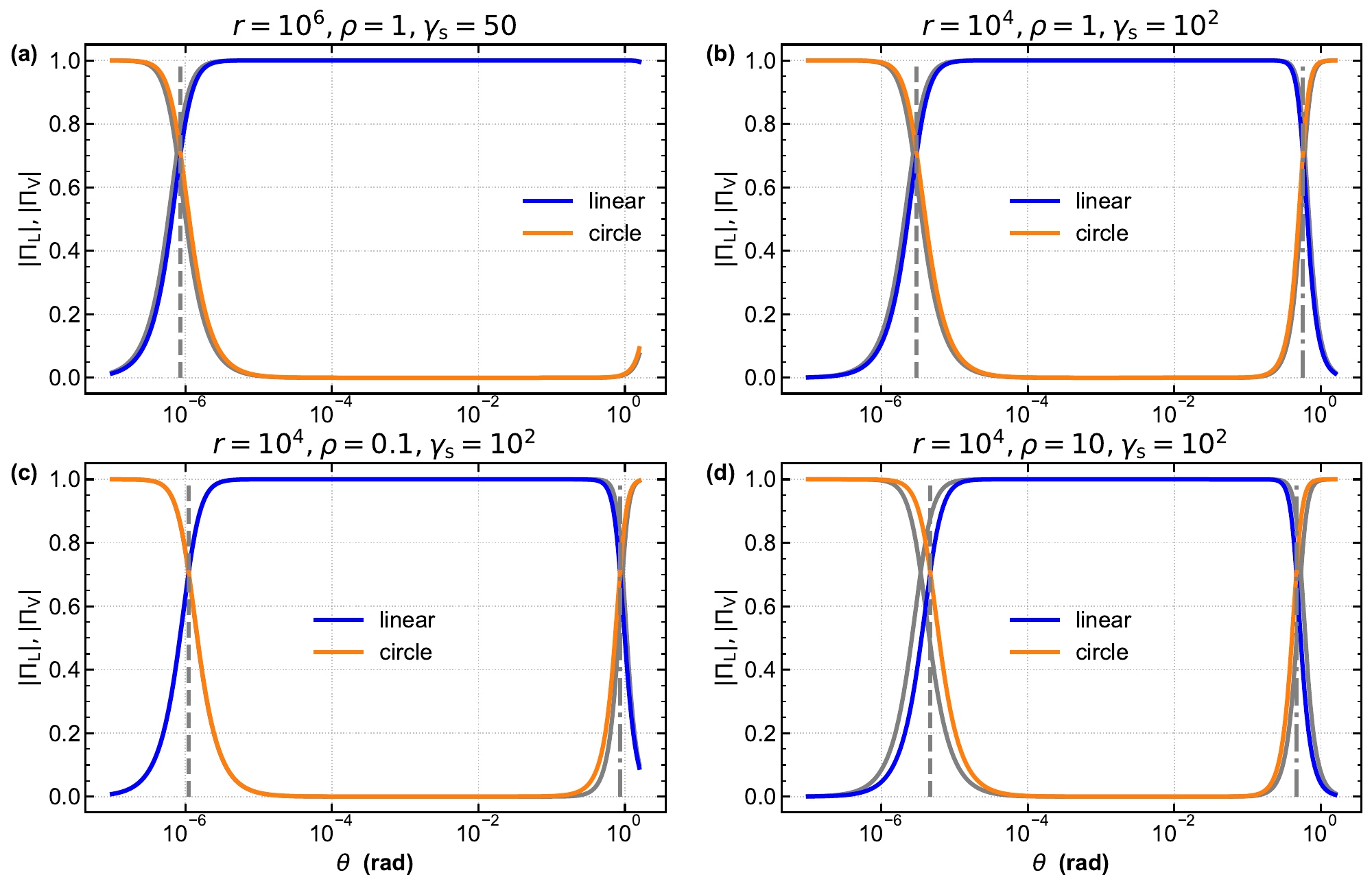}  
    \caption{The LP ($|\Pi_{\mathrm{L}}|$, blue lines) and CP ($|\Pi_{\mathrm{V}}|$, orange lines) degrees as a function of the angle between the wave vector and magnetic field for the intrinsically relativistic plasma. The transition angles $\theta_1$ and $\theta_2$ are given by numerical solutions of $|\mathcal{R}|=2$, which are represented by gray dashed and dotted-dashed vertical lines and their values are listed in Table \ref{tab:trtheta}. For $r=10^{6},\rho=1,\gamma_{\mathrm{s}}=50$, the second transition angle $\theta_2$ does not exist (panel (a)). For $r=10^{4},\gamma_{\mathrm{s}}=100$, transition angles of the case $\rho=1,0.1,10$ are shown in panel (b)-(d), respectively. The gray solid lines are the interpolation approximation \citep{Melrose2004}. In the ultra-relativistic case ($\rho \ll 1$), equation (\ref{eq:R_app}) is a good approximation.}
    \label{fig:Pi2}
\end{figure*}

First, we should estimate the radius at which magneto-ionic theory remains valid. In this work, we set the plasma rest frame as $K^{\prime}$ and the magnetar rest frame as $K$ in which the plasma is relativistic streaming with speed $v_{\mathrm{s}}=\beta_{\mathrm{s}} c$. The nonlinear or strong wave effect should be considered for the waves with extremely high luminosity \citep{Luan2014}. This effect can be described by a strong wave factor

\begin{equation} 
a_0=\frac{e E_0}{m_{\mathrm{e}}\omega c} \approx 2.3 \times 10^3  L_{\mathrm{iso}, 40}^{1 / 2} \nu_9^{-1} r_{0,9}^{-1},
\end{equation}
where $\nu$ and $E_0$ are the FRB's frequency and electric field strength, respectively. For repeating FRBs with narrow bandwidth $\Delta\nu$, the isotropic luminosity is 
\begin{equation}      
\begin{aligned} 
L_{\mathrm{iso}} &\simeq  4 \pi D_{\mathrm{L}}^2 S_{\nu, \mathrm{p}}\Delta\nu \\
& \simeq 1.3 \times 10^{40} \mathrm{erg} \mathrm{~s}^{-1}\left(\frac{D_{\mathrm{L}}}{10^{27} \mathrm{~cm}}\right)^2 \left(\frac{S_{\nu, \mathrm{p}}}{\mathrm{Jy}}\right)\left(\frac{\Delta \nu }{100~\mathrm{MHz}}\right),
\end{aligned}
\end{equation}
where $D_{\mathrm{L}}$ is the luminosity distance and $S_{\nu, \mathrm{p}}$ is the specific peak flux density. We focus on actively repeating FRBs with CP emission, such as FRB 20201124A and FRB 20220912A (see Table \ref{tab:CP_frb}). These sources exhibit isotropic luminosity in the range of $10^{38}-10^{41}$ erg s$^{-1}$, with a characteristic value of $\sim10^{40}$ erg s$^{-1}$, consistent with statistical findings reported in \cite{Wu2025}. Within the region with $1<\Omega_{\mathrm{e}}/\omega<a_0$, the particles will be accelerated to high Lorentz factors during resonance events with the waves. The study of the propagation of FRBs with high luminosity (e.g., $L_{\mathrm{iso}}\gtrsim10^{42}$ erg s$^{-1}$) in the magnetosphere can be found in \cite{Beloborodov2021,Beloborodov2022,Qu2022,Huang2024}. The linear magneto-ionic theory is only applicable when $\Omega_{\mathrm{e}}/\omega> a_0$, and the corresponding radius is

\begin{equation}  
r_0< R_2\simeq1.3\times 10^9 ~\mathrm{cm} ~B_{\mathrm{s}, 15}^{1 /2} L_{\mathrm{iso}, 40}^{-1 / 4}.
\end{equation}

In the magnetosphere, the waves have two orthogonal modes (X-mode and O-mode) when the wave frequency is much larger than the plasma frequency in the frame $K^\prime$, i.e., $\omega^{\prime}\gg\omega_\mathrm{p}^{\prime}$. The plasma frequency in the plasma rest frame is $\omega_\mathrm{p}^{\prime}=\sqrt{4 \pi e^2 \kappa n_{\mathrm{GJ}}^{\prime}/m_{\mathrm{e}}}=\sqrt{4 \pi e^2 \kappa n_{\mathrm{GJ}}/\gamma_{\mathrm{s}}m_{\mathrm{e}}}$. For simplicity, we only consider quasi-parallel propagation, so the wave frequency in the plasma rest frame is $\omega^{\prime}=\omega/2\gamma_{\mathrm{s}}$. From equation (\ref{eq:n_GJ}), we can get the limit of radius

\begin{equation}      
    r_0>R_1\simeq1.3\times 10^8 ~\mathrm{cm} ~\nu_9^{-2/3} \kappa^{1/3} \gamma_{\mathrm{s,2}}^{1/3}B_{\mathrm{s}, 15}^{1 /3}P^{-1 /3}.
\end{equation}

For typical active repeating FRB observation properties, the region such magneto-ionic theory is valid requires $R_2>R_1$, i.e.,
\begin{equation}     
    \kappa \gamma_{\mathrm{s}} \lesssim  10^5\nu_9^{2} B_{\mathrm{s}, 15}^{1 /2}L_{\mathrm{iso}, 40}^{-3/ 4}P. 
\end{equation}
The above conditions are easily achieved in the magnetar magnetosphere (e.g., $\gamma_{\mathrm{s}}\sim 10^2-10^3$ and $\kappa\sim 1-10^2$, \citealt{Daugherty1982,Arendt2002,Timokhin2015,Medin2010,Beloborodov2013}). In the region $R_1<r_0<R_2$, magnetoionic parameters satisfy approximately $X^\prime\ll1$ and $Y^\prime\gg1$, and equation (\ref{eq:R_cold}) becomes
\begin{equation}\label{eq:R_mag}  
\mathcal{R}^\prime\approx Y^\prime \sin ^2 \theta^\prime / |\eta| \cos \theta^\prime    
\end{equation}
in the plasma rest frame.

\begin{figure*}
    \centering
    \resizebox{\hsize}{!}{\includegraphics{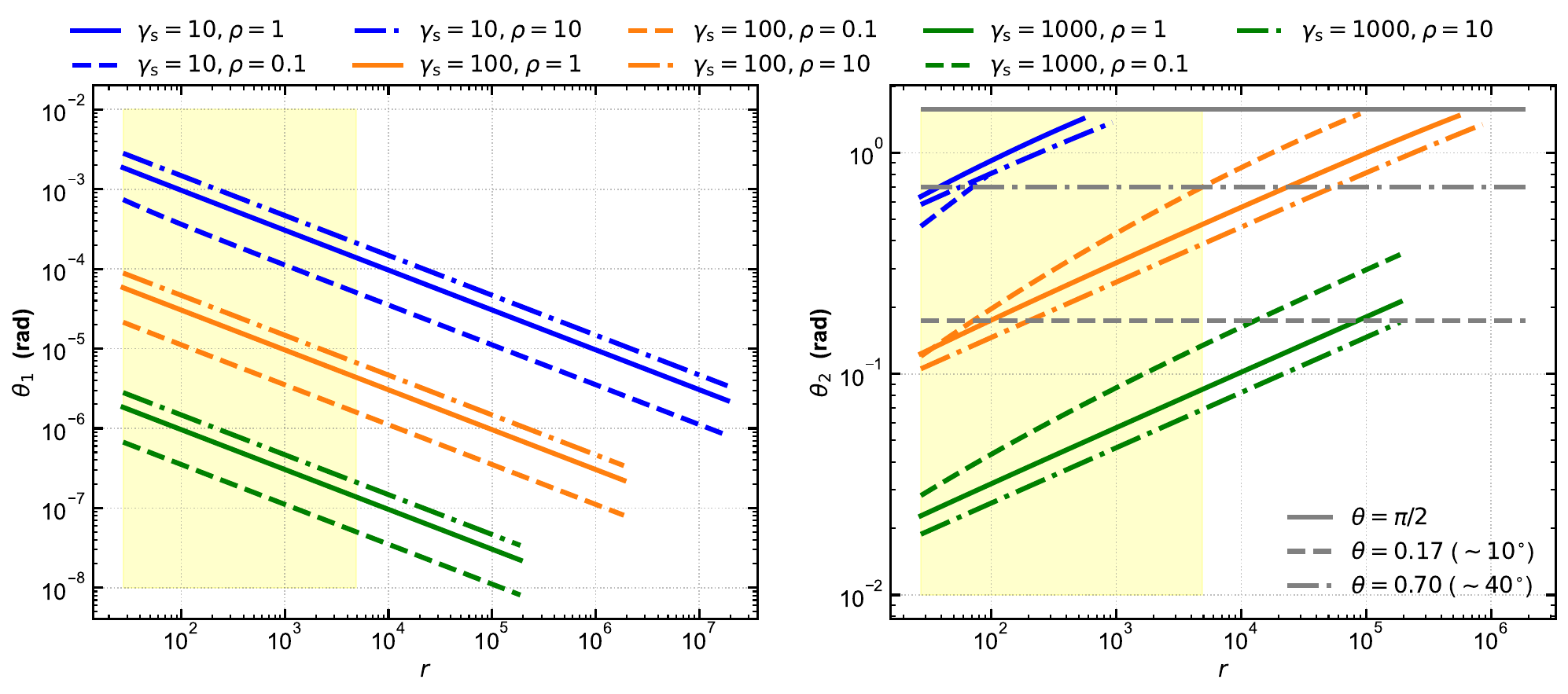}}
    \caption{Same as Figure \ref{fig:thetac1}, but for the intrinsically relativistic plasma.}
    \label{fig:thetac2}
\end{figure*}

\subsection{Relativistic streaming plasma}\label{sec:stream}

In the magnetosphere of a magnetar, charged particles nearly slide along magnetic field lines due to the strong magnetic field. The bulk relativistic motion of the plasma $\gamma_{\mathrm{s}}=\left(1-\beta_{\mathrm{s}}^2\right)^{-1 / 2}$ is set along the 3-axis. 
In the weak anisotropy approximation, the longitudinal part of the polarization is neglected. Thus, the polarization vector of the transverse part is also nearly transverse after the Lorentz transformation \citep{Melrose2004}. The axial ratio of the polarization ellipse $T$ depends on the parameter $\mathcal{R}$. After the Lorentz transformation, equation (\ref{eq:R_mag}) is expressed as
\begin{equation}\label{eq:R1} 
\mathcal{R} \approx \frac{r \sin ^2 \theta}{\gamma_{\mathrm{s}}^3\left(1- \beta_{\mathrm{s}} \cos \theta\right)^2\left( \cos \theta-\beta_{\mathrm{s}}\right)} 
\end{equation}
in the magnetar rest frame. For $\beta_{\mathrm{s}} \approx 1-1 / 2 \gamma_{\mathrm{s}}^2$ and $\gamma_{\mathrm{s}}^2 \theta^2 \ll 1$, the parameter $\mathcal{R}$ can be approximated as
\begin{equation}
\begin{aligned}
    \mathcal{R}&\approx \frac{8 r\gamma_{\mathrm{s}}^3 \theta^2}{\left(1+\gamma_{\mathrm{s}}^2 \theta^2\right)^2} \frac{1}{1-\gamma_{\mathrm{s}}^2 \theta^ 2}\\
    &\approx 8r\gamma_{\mathrm{s}}^3\theta^2
\end{aligned}
\end{equation}
The first transition angle ($|\mathcal{R}|=2$) is 
\begin{equation}
\theta_{\mathrm{1}}\approx\sqrt{1/r\gamma_{\mathrm{s}}^3}/2=5\times10^{-6}~\mathrm{rad}~r_{4}^{-1/2}\gamma_{\mathrm{s},2}^{-3/2}.
\end{equation}

\begin{figure*}
    \centering
    \resizebox{\hsize}{!}{\includegraphics{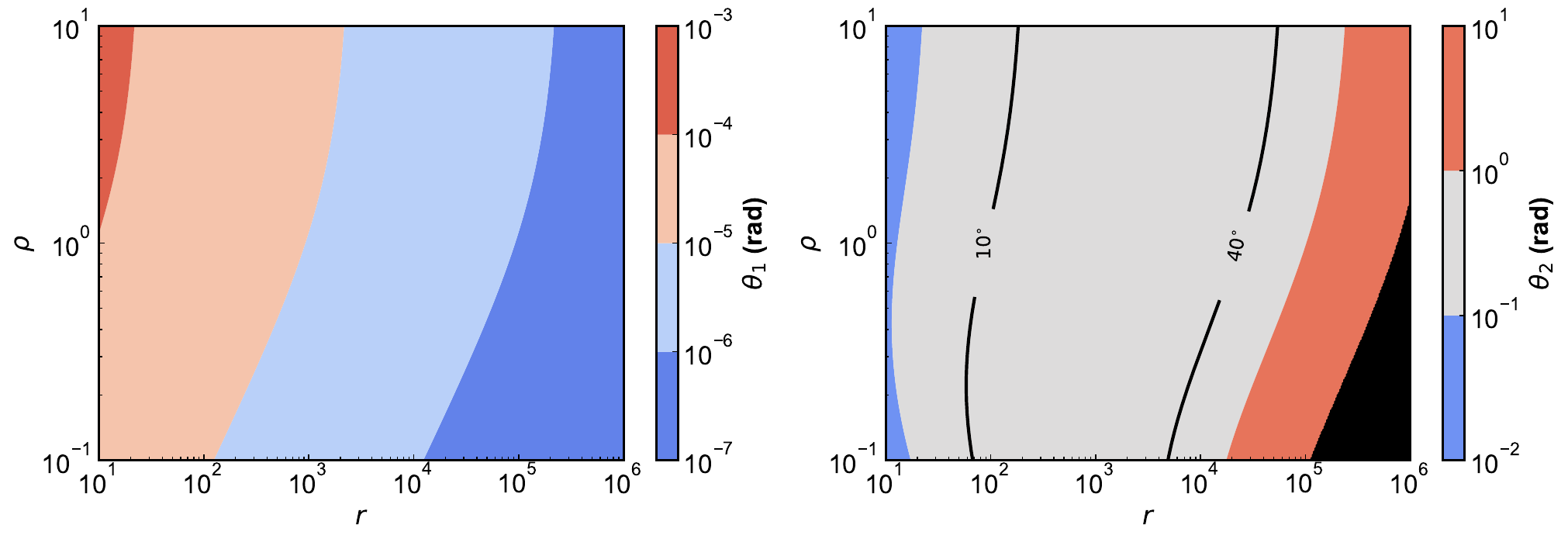}}
    \caption{Contours of transition angles for different inverse temperatures $\rho$ and $r$. The bulk Lorentz factor is fixed at $\gamma_{\mathrm{s}}=100$. The first transition angle is about $\theta_1\sim 10^{-7}-10^{-3}$ rad, and the second transition angle is about $\theta_1\sim 10^{-2}-1$ rad. The limits of the maximum angles $\theta_{\max}=10^{\circ}$, and $\theta_{\max}=40^{\circ}$ are also shown in black lines and the black region represents the second transition angle that does not exist for $\theta<\pi/2$. }
    \label{fig:theta_i12}
\end{figure*}

The Lorentz transformation shrinks the forward cone but expands the backward cone into the forward hemisphere in the frame $K$, making it possible to have two transition angles instead of just one in non-relativistic cold plasma for forward propagation (see Figure \ref{fig:trtheta}). The high CP caused by this effect is called aberrated backward circular polarization (ABCP, \citealt{Melrose2004}). For $\theta^\prime=\pi / 2$ in the plasma rest frame, the corresponding angle is $\theta\sim 1/\gamma_{\mathrm{s}}$ in the magnetar rest frame.
The second transition angle $1/\gamma_{\mathrm{s}}<\theta_2<\pi / 2$ exists when $|\mathcal{R}| \lesssim 2$ for $\theta=\pi / 2$, i.e.,
\begin{equation}\label{eq:theta2_cold}
    r=\frac{\Omega_\mathrm{e}}{\omega|\eta|}\lesssim 2\gamma_\mathrm{s}^3\sim 2\times10^6 \gamma_{\mathrm{s},2}^{3}.
\end{equation}

When equation (\ref{eq:theta2_cold}) is satisfied, $\theta_2$ can be obtained by sovling $|\mathcal{R}| = 2$ numerically. The CP and LP degree as a function of the angle between the wave vector and magnetic field is shown in Figure \ref{fig:Pi1}. The transition angles of different parameters are listed in Table \ref{tab:trtheta}. For a strong background magnetic field and low Lorentz factor (e.g., $r=10^{6},\gamma_{\mathrm{s}}=50$), the second transition angle $\theta_2$ does not exist (see panel (a) in Figure \ref{fig:Pi1}). For a weak background magnetic field and large Lorentz factor (e.g., $r=10^{4},\gamma_{\mathrm{s}}=100$), there are two transition angles (see panel (b) in Figure \ref{fig:Pi1}). The transition angles $\theta_1$ and $\theta_2$ are represented by gray dashed and dotted-dashed vertical lines, respectively. 

Here, we estimate the range of the parameter $r$ based on some typical magnetospheric parameters and assumptions commonly used in FRB models. In some magnetospheric emission models, FRBs are assumed to be generated in open magnetic field lines and the radiation position is $R_{\mathrm{FRB}}\sim100R_{\star}$ \citep{Yang2018,Kumar2020,Lu2020,Zhang2022,Liu2023}. If $R_{\mathrm{FRB}}<R_1$, the polarization properties may change as the FRB passes through the region $R_1<r_0<R_2$. The maximum value of $r$ corresponds to the plasma at the radius $R_1$
\begin{equation}
    r_{\max}=\frac{\Omega_\mathrm{e}(R_1)}{\omega|\eta|}\simeq1.25\times10^6\nu_9\gamma_{\mathrm{s,2}}^{-1}P.
\end{equation}
The value of $r$ reaches its minimum value at the radius $R_2$ and $|\eta|\sim1$
\begin{equation}
r_{\min}=\frac{\Omega_\mathrm{e}(R_2)}{\omega|\eta|}\simeq1.3\times10^3\nu_9^{-1}B_{\mathrm{s}, 15}^{-1 /2} L_{\mathrm{iso}, 40}^{3/ 4}.    
\end{equation}

Figure \ref{fig:thetac1} shows the transition angles $\theta_1$ and $\theta_2$ for a wide range of $r$. The characteristic frequency $\nu\sim1.5$ GHz is used. The shade region represents the value of $r_{\min}$ for FRBs with the isotropic luminosity of $10^{38}-10^{41}$ erg s$^{-1}$.

\subsection{Intrinsically relativistic  plasma}\label{sec:intrinsic}
In section \ref{sec:stream}, the plasma is assumed to be relativistic streaming, but the random motions of the particles in the plasma rest frame are ignored. 
From the numerical simulations of the pair creation, the pair plasma in the magnetosphere can be described in a relativistic thermal distribution \citep{Hibschman2001,Arendt2002}. The inverse temperature of the background plasma is $\rho\sim1$ \citep{Arendt2002,Lopez2015,Benacek2021}, where $\rho=m_{\mathrm{e}}c^2/k_{\mathrm{B}}T$ with $k_{\mathrm{B}}$ being the Boltzmann constant. The 1-D Jüttner distribution is taken in this work
\begin{equation}\label{eq:gu}
g(u)=\frac{n_{\mathrm{e}} e^{-\rho \gamma}}{2 K_1(\rho)}, 
\end{equation}
where $u=\gamma \beta$ is the four-velocity. The distribution function is normalized to the total electron number density $\int_{-\infty}^{\infty} \mathrm{d} u g(u)=n_{\mathrm{e}}=n_{+}+n_{-}$, and $K_1(\rho)$ is the modified Bessel function of one type. The wave equation in the magnetosphere plasma of a magnetar can be obtained from the dielectric tensor of the relativistic pair plasma (see Appendix \ref{sec:appendixB}).

In the weak anisotropy approximation, one has $z^\prime=\omega^\prime /k^\prime c \cos \theta^\prime \rightarrow \sec \theta^\prime$ \citep{Melrose2004}. 
Similar to the cold plasma, the polarization ellipse in the frame $K^\prime$ can also be obtained from the relativistic dispersion relation\footnote{The more accurate results can be found in \cite{Rafat2019}. However, the results of \cite{Melrose2004} are still applicable under the typical parameters in the magnetosphere (see Appendix \ref{sec:appendixB}).} \citep{Melrose2004}. In the frame $K^\prime$, the parameter $\mathcal{R}^\prime$ is
\begin{equation}
\begin{aligned}
\mathcal{R}^{\prime}&=\frac{\left(\tilde{P} \tilde{S}-\tilde{S}^2+\tilde{D}^2\right) \sin ^2 \theta^{\prime}}{\tilde{P} \tilde{D} \cos \theta^{\prime}}\\
    &=\frac{\Omega_{\mathrm{e}}}{\eta \omega^{\prime}} \frac{\sin ^2 \theta^{\prime}}{\cos \theta^{\prime}} \sec ^2 \theta^{\prime} W(\sec \theta^{\prime}).
\end{aligned}
\end{equation}
The parameter $\mathcal{R}$ in the magnetar rest frame is 
\begin{equation}\label{eq:R2} 
    \mathcal{R} \approx \frac{r \sin ^2 \theta}{\gamma_{\mathrm{s}}^3\left( \cos \theta-\beta_{\mathrm{s}}\right)^3}W\left(\frac{1-\beta_{\mathrm{s}}\cos \theta}{\cos \theta-\beta_{\mathrm{s}}}  \right).
\end{equation}

The LP and CP degrees for the intrinsically relativistic plasma are shown in Figure \ref{fig:Pi2}. The transition angles $\theta_1$ and $\theta_2$ are given by numerical solutions of $|\mathcal{R}|=2$, which are represented by gray dashed and dotted-dashed vertical lines. Their values are listed in Table \ref{tab:trtheta}. For $r=10^{6},\rho=1,\gamma_{\mathrm{s}}=50$, the second transition angle $\theta_2$ does not exist (panel (a)). For $r=10^{4},\gamma_{\mathrm{s}}=100$, transition angles of the case $\rho=1,0.1,10$ are shown in panel (b)-(d), respectively. The gray solid lines are the interpolation approximation taken from \cite{Melrose2004}
\begin{equation}\label{eq:R_app}
\mathcal{R}  \approx \begin{cases}16r\langle\gamma\rangle \gamma_{\mathrm{s}}^3 \theta^{2} & \text { for } \theta \lesssim \theta_{0+}, \\ 
2r \gamma_{\mathrm{s}}\left\langle\frac{1}{\gamma^3}\right\rangle \frac{1+\gamma_{\mathrm{s}}^4 \theta^{4}}{\gamma_{\mathrm{s}}^2 \theta^{2}\left(1-\gamma_{\mathrm{s}}^2 \theta^{2}\right)} & \text { for }\theta_{0-} \lesssim \theta \lesssim \theta_{0+},\\
r\frac{\langle\gamma\rangle \cos ^2\left(\theta / 2\right)}{\gamma_{\mathrm{s}}^3 \sin ^4\left(\theta / 2\right)} & \text { for } \theta \gtrsim \theta_{0-},\end{cases}
\end{equation}
where $\theta_{0+}=\theta_0/2\gamma_{\mathrm{s}},\theta_{0-}=2/\gamma_{\mathrm{s}} \theta_0$ and $\theta_0 \sim 1 /\langle\gamma\rangle^{1 / 2}$. In the ultra-relativistic case ($\rho \ll 1$), equation (\ref{eq:R_app}) is a good 
approximation. For $\rho = 10$, the average Lorentz factor is $\langle\gamma\rangle\sim1.05$, and the transition angles are close to the cold plasma case (see Table \ref{tab:trtheta}). The ABCP effect can be described in $\theta \lesssim \theta_{0+}$ and $\theta \gtrsim \theta_{0-}$. For $\theta_{0-} \lesssim \theta \lesssim \theta_{0+}$, possible CP is caused by the relativistic particle distribution, named intrinsically relativistic CP (IRCP, \citealt{Melrose2004}).  The relativistic dispersion relation involves the average over the distribution, which can be given by \citep{MG99}
\begin{align}
&\langle\gamma\rangle =\frac{K_2(\rho)+K_0(\rho)}{2 K_1(\rho)} , \\
&\langle1/\gamma^{n+1}\rangle =\frac{K i_n(\rho)}{K_1(\rho)},
\end{align}
where $K_n(\rho)$ are modified Bessel functions the $n$th kind, and $K i_n(\rho)$ are Bickley functions.

Transition angles can be estimated from the approximation results for $\rho \ll 1$. From the equation (\ref{eq:R_app}), first transition angle is
\begin{equation}
\theta_{\mathrm{1}}\approx\sqrt{1/8r\langle\gamma\rangle\gamma_{\mathrm{s}}^3}=1.1\times10^{-6}~\mathrm{rad}~r_{4}^{-1/2}\langle\gamma\rangle_1^{-1/2}\gamma_{\mathrm{s},2}^{-3/2}.
\end{equation}
The second transition angle $\theta_{\mathrm{2}}$ exists when $|\mathcal{R}| \lesssim 2$ for $\theta=\pi / 2$ in equation (\ref{eq:R_app}), i.e.,
\begin{equation}
    r=\frac{\Omega_\mathrm{e}}{\omega|\eta|}\lesssim \gamma_\mathrm{s}^3\langle\gamma\rangle^{-1}\sim 10^5 \gamma_{\mathrm{s},2}^{3}\langle\gamma\rangle_{1}^{-1}.
\end{equation}

Transition angles for $r\sim10-10^6$ are shown in Figure \ref{fig:thetac2}. Contours of transition angles for different inverse temperatures $\rho$ and $r$ are shown in Figure \ref{fig:theta_i12}. The bulk Lorentz factor is fixed at $\gamma_{\mathrm{s}}=100$. The first transition angle is about $\theta_1\sim 10^{-7}-10^{-3}$ rad, and the second transition angle is about $\theta_1\sim 10^{-2}-1$ rad. The limits of the maximum angles $\theta_{\max}=10^{\circ}$, and $\theta_{\max}=40^{\circ}$ are also shown in black lines in Figure \ref{fig:theta_i12} and the black region represents the second transition angle that does not exist for $\theta<\pi/2$. 

High CP for IRCP requires $|\mathcal{R}|\lesssim 2$ for $\theta_{0-} \lesssim \theta \lesssim \theta_{0+}$, which can be further simplified to $2r \gamma_{\mathrm{s}}\left\langle1/\gamma^3\right\rangle \lesssim 2$ for $\gamma_{\mathrm{s}}^2 \theta^{2}\gg1$. For $\rho=0.1$, the value of $\left\langle1/\gamma^3\right\rangle$ has the same order of magnitude as $1/\left\langle\gamma\right\rangle\sim0.1$. The required frequency 
\begin{equation}
    \nu\gtrsim 250~\mathrm{GHz}~B_\mathrm{s, 15}P^{-3}R_{\star,6}^3\gamma_{\mathrm{s,2}}\left\langle\gamma\right\rangle_1^{-1}|\eta|^{-1}
\end{equation}
is too high even near the light cylinder. However, for our interested $\sim$GHz band, the effects of IRCP can contribute a low degree of CP
\begin{equation}\label{eq:cp_ircp}    
    \Pi_{\mathrm{V}}\approx\frac{2}{|\mathcal{R}|}\sim 10\% ~|r|_1^{-1}\gamma_{\mathrm{s,1}}^{-1}\left\langle\gamma\right\rangle_1.
\end{equation}

\begin{table*}
    \centering
    \caption{The CP measurements of active repeating FRBs.}
    \label{tab:CP_frb}
    \begin{threeparttable}
        \begin{tabular}{lccccccccc}
            \hline
            Source &  Burst No. & $|\Pi_{\mathrm{V}}|>20\%$ & $|\Pi_{\mathrm{V}}|>50\%$ & $F_{20}$ & $F_{50}$ & $\Pi_{\mathrm{V,+}}$ & $\Pi_{\mathrm{V,-}}$ &$N_{+}:N_{-}$ & Ref. \\
            \hline
            FRB 20211124A & 1103 & 136 & 12 & 0.12 & 0.01 & &  &  & [1]\\
            FRB 20211124A & 536 & 106 & 15 & 0.20 & 0.03 & 270 & 266 & 1.02 & [2,3] \\
            FRB 20220912A & 1076 & 150 & 16 & 0.14 & 0.01 & 570 & 500 & 1.14 &[4] \\
            FRB 20220912A & 128 & 66 & 7 & 0.51 & 0.05 & 101 & 27 & 3.74 &[5] \\
            FRB 20240114A & 297 & 42 & 1 & 0.14 & 0.003 & 141 & 154 & 0.92 & [6] \\
            \hline
\end{tabular}
    \begin{tablenotes}
        \item Note: Column (1): Source; Column (2): Total bursts number with polarization measurements; Column (3): Bursts number with $|\Pi_{\mathrm{V}}|>20\%$; Column (4): Bursts number with $|\Pi_{\mathrm{V}}|>50\%$; Column (5): The proportion of bursts with $|\Pi_{\mathrm{V}}|>20\%$; Column (6): The proportion of bursts with $|\Pi_{\mathrm{V}}|>50\%$; Column (7): Bursts number with $\Pi_{\mathrm{V}}>0$; Column (8): Bursts number with $\Pi_{\mathrm{V}}<0$; Column (9): The ratio of the bursts number of $\Pi_{\mathrm{V}}>0$ and $\Pi_{\mathrm{V}}<0$; Column (10): References: [1] \cite{Xu2022}; [2] \cite{Jiang2022}; [3] \cite{Jiang2024}; [4] \cite{ZYK2023}; [5] \cite{Feng2024}; [6] \cite{Xie2025}.
    \end{tablenotes}
    \end{threeparttable}
\end{table*}

\begin{figure}
    \centering
\includegraphics[width=1\linewidth]{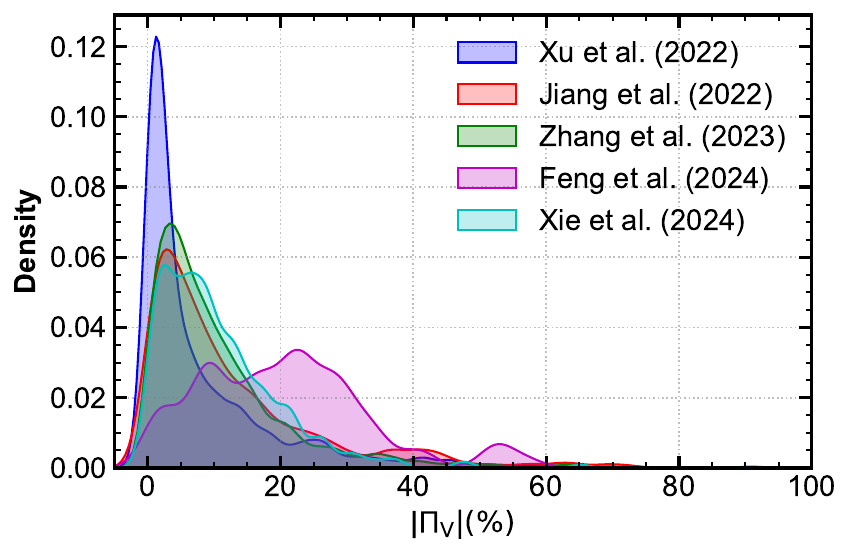}
    \caption{The kernel density estimation (KDE) of $|\Pi_{\mathrm{V}}|$ from some active repeating FRBs, e.g., FRB 20211124A from FAST \citep{Xu2022,Jiang2022,Jiang2024}, FRB 20220912A from FAST \citep{ZYK2023} and GBT \citep{Feng2024} and FRB 20240114A from GBT \citep{Xie2025}.}
    \label{fig:V_abs}
\end{figure}

\begin{figure}
    \centering
    \includegraphics[width = 1\linewidth]{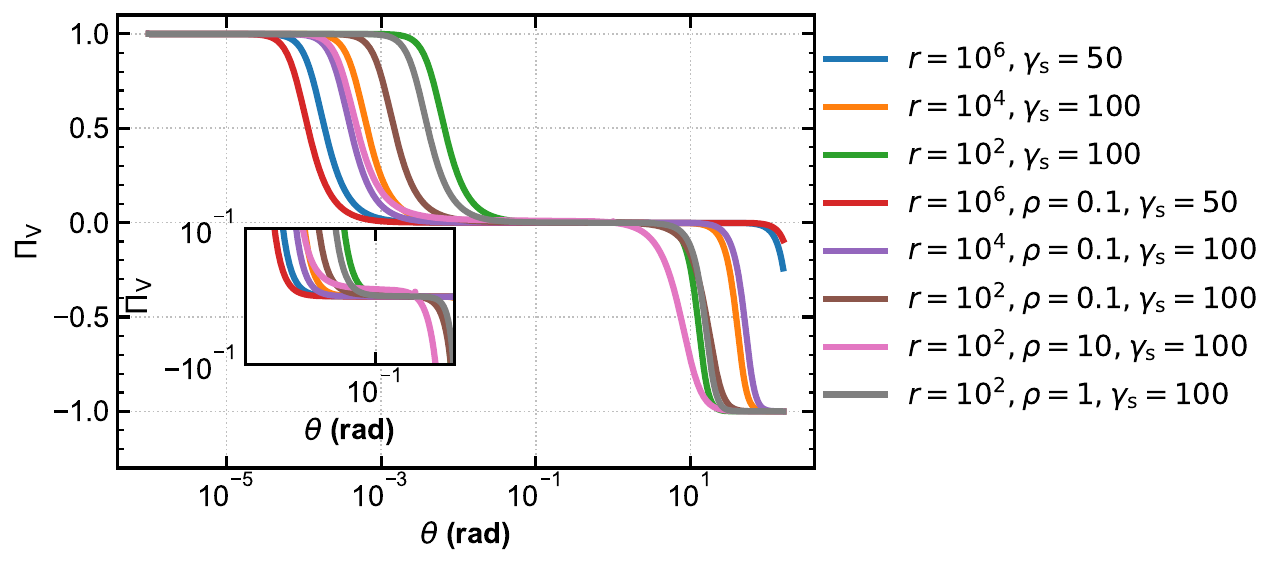}
    \caption{The CP degrees as a function of angle for different plasma models in the magnetosphere of a magnetar. The handedness of CP is opposite for $\theta<1/\gamma_{\mathrm{s}}$ and $\theta>1/\gamma_{\mathrm{s}}$ for the plasma with relativistic motion $\gamma_{\mathrm{s}}$ (see Figure \ref{fig:trtheta}(b)).}
    \label{fig:rlcp}
\end{figure}

\section{Comparison to observations}\label{sec:obs}

Before comparing with specific observed FRB polarization data, we present the propagation of FRBs in magnetar magnetospheres, which creates additional constraints on propagation-induced CP due to magnetar magnetospheric parameters or FRB emission. If FRBs emit along the tangent of the field line, the propagation angle between the wave vector and the magnetic field can be constrained by the dipolar field geometry. \cite{Qu2022} found that if the emission radius is $\lesssim 10^8 \mathrm{~cm}$, the value of $\theta$ is smaller than $10^{\circ}$ for the most case. When the wave propagates to the light cylinder, $\theta$ reaches its maximum which is $\theta_{\max}\sim40^{\circ}$ \citep{Qu2022}. The limits of the maximum angles $\theta_{\max}=90^{\circ}$, $\theta_{\max}=10^{\circ}(\sim0.17~\mathrm{rad})$, and $\theta_{\max}=40^{\circ}(\sim0.70~\mathrm{rad})$ are also shown in Figures \ref{fig:thetac1} and \ref{fig:thetac2} as gray horizontal solid, dashed, and dotted-dashed lines. Under certain parameters, the second transition angle $\theta_2$ approaches or exceeds the maximum angle. Combined with the fact that the first transition angle $\theta_1$ is so small for $\gamma_{\mathrm{s}}\gg1$, makes high CP extremely rare.

Here we collect the CP measurement data of some active repeating FRBs, e.g., FRB 20211124A from FAST \citep{Xu2022,Jiang2022,Jiang2024}, FRB 20220912A from FAST \citep{ZYK2023} and GBT \citep{Feng2024} and FRB 20240114A from GBT \citep{Xie2025}. The polarization properties are listed in Table \ref{tab:CP_frb} and the kernel density estimation (KDE) of $|\Pi_{\mathrm{V}}|$ is shown in Figure \ref{fig:V_abs}. Except for the GBT dataset of FRB 20220912A \citep{Feng2024}, in other samples, only about $10\%-20\%$ of the bursts have CP above 20\%. However, the total number of bursts in the GBT datasets of FRB 20220912A is only over one hundred \citep{Feng2024}, which is almost one-tenth of the FAST dataset \citep{ZYK2023}. Therefore, in terms of the total number of bursts of FRB 20220912A, high CP is still rare. For bursts with CP greater than 50\%, they account for $0.3\%-5\%$ of all bursts with polarization measurements.

\begin{figure*}
    \centering
    \resizebox{\hsize}{!}{\includegraphics{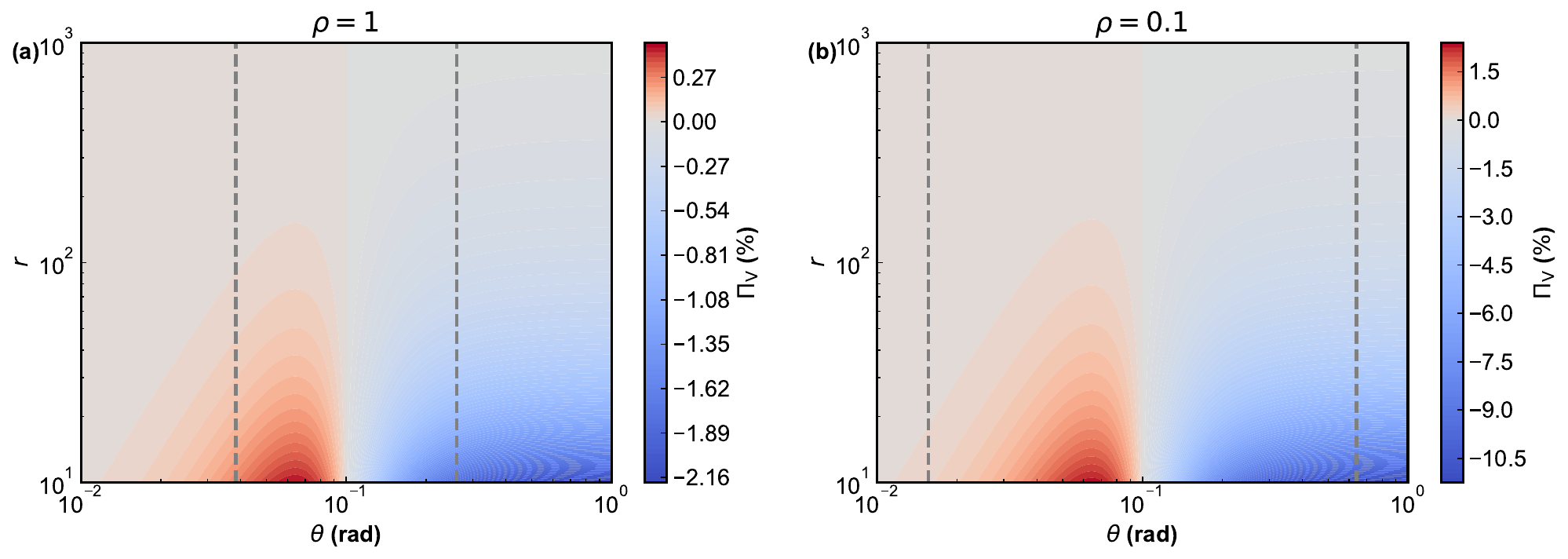}}
    \caption{The CP contours caused by the IRCP effect in $|r|-\theta$ plane for $\eta=0.1$ and $\gamma_{\mathrm{s}}=10$. The two grey vertical dashed lines represent $\theta_{0-}$ and $\theta_{0+}$ respectively. The intrinsically relativistic distribution enhances the CP for $\theta_{0-} \lesssim \theta \lesssim \theta_{0+}$, which is very different from cold plasma.}
    \label{fig:inc_cp}
\end{figure*}

The required wave frequency for high CP caused by IRCP is too high for FRBs, but IRCP can explain some bursts with low CP (equation (\ref{eq:cp_ircp})). The CP contours caused by the IRCP effect in the $r-\theta$ plane for $\gamma_{\mathrm{s}}=10$ are shown in Figure \ref{fig:inc_cp}. The two grey vertical dashed lines represent $\theta_{0-}$ and $\theta_{0+}$ respectively. 
IRCP can account for a small degree of CP with random handedness from pulse to pulse. The CP degrees as a function of angle for different plasma models in the magnetosphere of a magnetar are shown in Figure \ref{fig:rlcp}. 

\begin{figure}
    \centering
\includegraphics[width=1\linewidth]{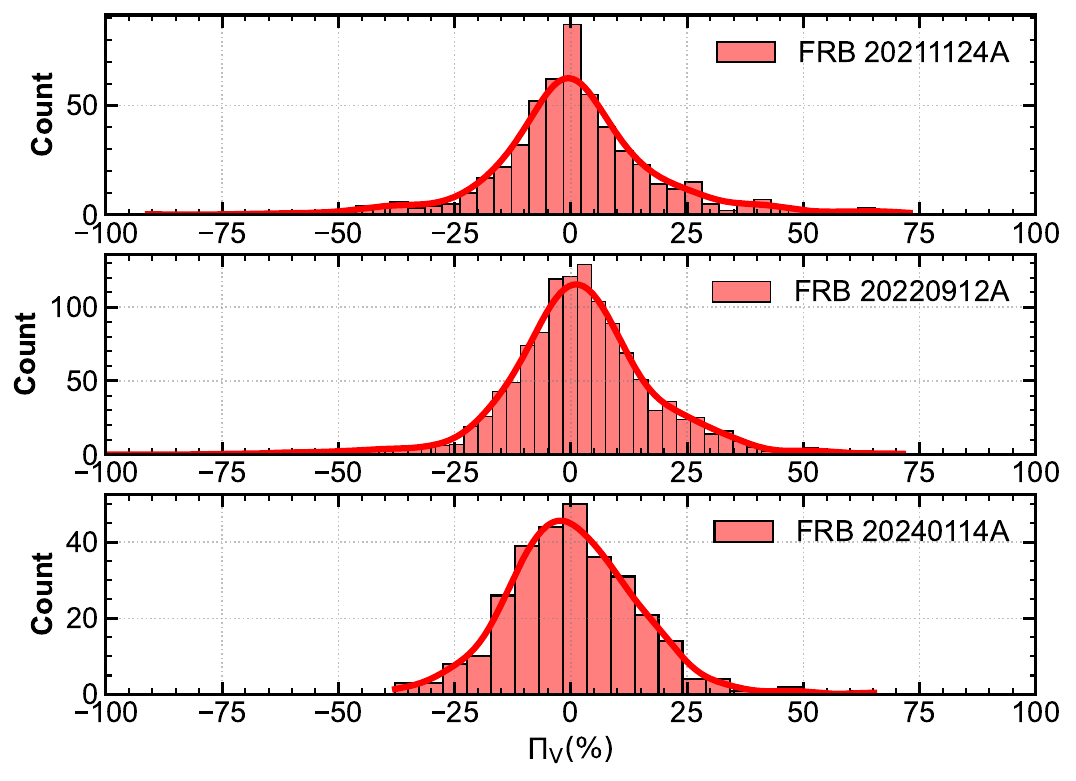}
    \caption{The CP distributions with different handedness of FRB 20201124A \citep{Jiang2022,Jiang2024}, FRB 20220912A \citep{ZYK2023} and 20240114A \citep{Xie2025}.}
    \label{fig:V_frb}
\end{figure}

The handedness of CP is opposite for $\theta<1/\gamma_{\mathrm{s}}$ and $\theta>1/\gamma_{\mathrm{s}}$ for the plasma with relativistic motion (see Figure \ref{fig:trtheta}(b)), which is a natural extension of the non-relativistic case. The CP distributions with different handedness of FRB 20201124A \citep{Jiang2022,Jiang2024}, FRB 20220912A \citep{ZYK2023} and 20240114A \citep{Xie2025} are shown in Figure \ref{fig:V_frb}. Table \ref{tab:CP_frb} shows that about 80\%-90\% of the bursts have a polarization degree below 20\%. The number of left-handed and right-handed circularly polarized bursts can be found in Table \ref{tab:CP_frb}, and the radio of them is close to 1:1. The small degree of CP with random handedness is consistent with the case of $\theta_{0-} \lesssim \theta \lesssim \theta_{0+}$ (see equation (\ref{eq:R_app})). From Figure \ref{fig:V_frb}, there are also different handedness for bursts with high circular polarization, which is rare but exists. Recently, extreme high CP of $(73.5 \pm 0.7) \%$ and $(-90.9 \pm 1.1) \%$ has been detected in FRB 20201124A \citep{Jiang2024}. The high CP of different handedness originates from $\theta \lesssim \theta_{0+}$ and $\theta \gtrsim \theta_{0-}$ (see equation (\ref{eq:R_app})).

\begin{figure*}
    \centering
\resizebox{\hsize}{!}{\includegraphics{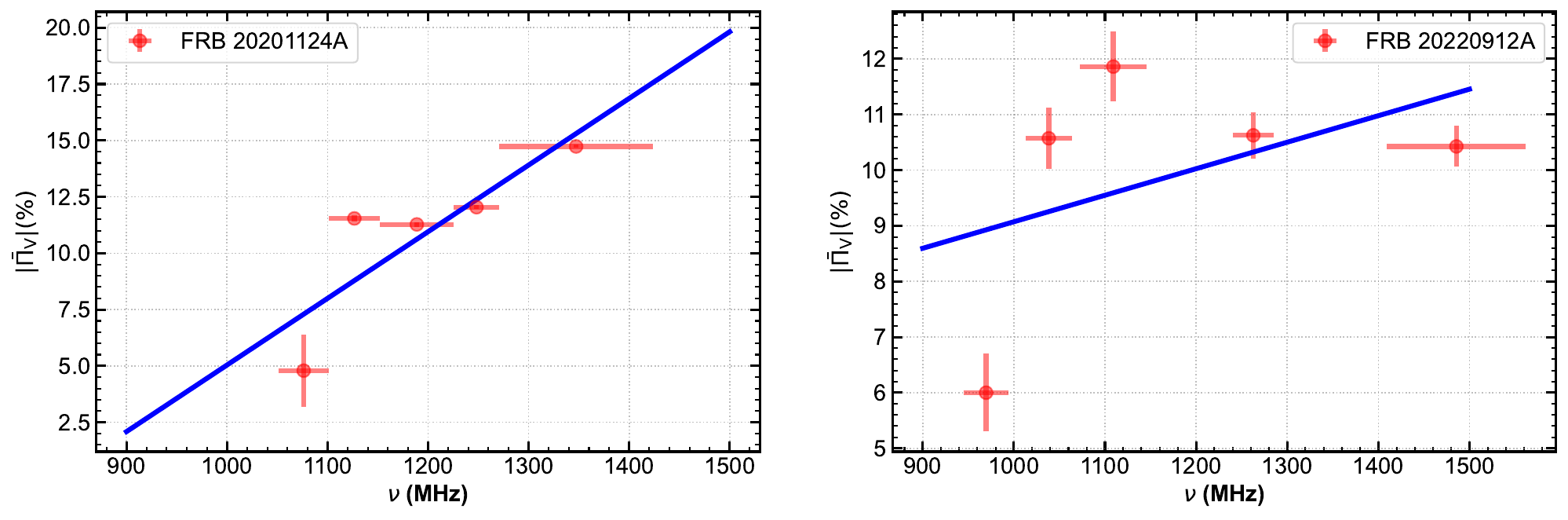}}
    \caption{The frequency-dependent averaged CP degree of FRB 20211124A \citep{Jiang2022,Jiang2024} and FRB 20220912A \citep{ZYK2023}. The red dots represent the binned data divided equally by the burst number, and the blue solid line is the result of the linear fit.}
    \label{fig:V_nu}
\end{figure*}

From equations (\ref{eq:R1}) and (\ref{eq:R2}), we can see that $\mathcal{R}  \propto r\propto\omega^{-1}$. The low degree of CP can be estimated as $\Pi_{\mathrm{V}}\approx2/\mathcal{R}$, which means that the CP degree should increase (linearly) with frequency. \cite{Petrova2000} have found similar results, and they suggest that the increase of $|\Pi_{\mathrm{V}}|$ with frequency is due to the more obvious refraction at high frequencies. Therefore, we can test whether propagation effects in magnetospheric  plasma play a role by studying the frequency dependence of the CP degree. Considering that the GBT observations of FRB 20240114A are only concentrated in a very narrow band of 720–920 MHz \citep{Xie2025}, we only used the FAST L-band (1.0-1.5 GHz) data of FRB 20211124A \citep{Jiang2024} and FRB 20220912A \citep{ZYK2023} to study the frequency dependence of CP. The averaged CP degree of FRB 20211124A \citep{Jiang2022,Jiang2024} and FRB 20220912A \citep{ZYK2023} in different frequency bins divided equally by the burst number are shown in red dots in Figure \ref{fig:V_nu}. The blue solid lines are the results of the linear fit and the slopes of the best fit for FRB 20211124A and FRB 20220912A are 0.3 GHz$^{-1}$ and 0.05 GHz$^{-1}$, respectively. For FRB 20211124A and FRB 20220912A, the average CP degree shows an increasing trend with frequency.

\section{Conclusions}\label{sec:con}
Recently, polarization properties have been found to be able to change over short periods of time, which disfavors the out-of-magnetosphere emission and propagation models \citep{Jiang2024}. Some magnetospheric radiation models (e.g., coherent CR or ICS) predict that CP, produced by off-axis emission, will be systematically fainter, which is also inconsistent with the latest observations \citep{Jiang2024}. In this work, we interpret the CP as propagation effects caused by the relativistic plasma in the magnetosphere, which were proposed to explain the observed CP in radio pulsars \citep{Melrose2004}.

For cold plasma, polarization properties depend on the transition angle $\theta_{\mathrm{c}}$: high LP for $\theta_{\mathrm{c}} \lesssim \theta \lesssim \pi-\theta_{\mathrm{c}}$, high CP with different handedness for $\theta \lesssim \theta_{\mathrm{c}}$ and $\pi-\theta_{\mathrm{c}} \lesssim \theta \leqslant \pi$. For relativistic plasma, the Lorentz transformation and the relativistic particle distribution should be taken into consideration, and specific effects are as follows:
\begin{itemize}
    \item For the relativistically streaming plasma, there are two transition angles for $0<\theta<\pi/2$ in the magnetar rest frame because the Lorentz transformation expands the backward cone into the forward hemisphere. CP caused by this effect is referred as aberrated backward circular polarization (ABCP).
    \item From the study of the pair creation, the plasma in the magnetosphere can be described in a relativistic thermal distribution \citep{Hibschman2001,Arendt2002}. CP caused by the relativistic particle distribution is named intrinsically relativistic circular polarization (IRCP).
    \item High CP for the relativistically streaming plasma requires $\theta<\theta_1$ or $\theta_2<\theta<\pi/2$. However, $\theta_1$ is  small for $\gamma_{\mathrm{s}}\gg1$ and $\theta_2$ is limited by the maximum propagation angle, making high CP rare. 
    \item The required wave frequency for high CP caused by IRCP is too high for FRBs, but IRCP can explain some bursts with low CP.
    \item The handedness of CP is opposite for $\theta<1/\gamma_{\mathrm{s}}$ and $\theta>1/\gamma_{\mathrm{s}}$ for relativistic streaming plasma with the bulk Lorentz factor $\gamma_{\mathrm{s}}$, which is consistent the observations of repeating FRBs.
    \item We also found that the average CP degree of FRB 20211124A \citep{Jiang2024} and FRB 20220912A \citep{ZYK2023} increases with frequency in the L-band of FAST, which is consistent with theoretical expectations of propagation effects in magnetospheric plasma \citep{Melrose2004,Petrova2000}.
    \end{itemize}

\begin{acknowledgements}
We thank the anonymous referee for helpful comments. We thank Q. Wu and Y. H. Qu for helpful discussions and J. C. Jiang for providing the data of FRB 20201124A. This work made use of data from the FAST. FAST is a Chinese national megascience facility, built and operated by the National Astronomical Observatories, Chinese Academy of Sciences. This work was supported by the National Natural Science Foundation of China (grant Nos. 12494575 and 12273009).

\end{acknowledgements}

%
\bibliographystyle{aa} 
\bibliography{ms} 
%

\begin{appendix}
\section{Wave dispersion in cold plasma}\label{sec:appendixA}
\subsection{Dielectric tensor and dispersion relation for cold plasma}\label{app:cold}

In cold plasma, the dielectric tensor $K_{i j}$ is \citep{Stix1962}
\begin{equation}\label{eq:Kij_cold}
K_{i j}=\left(\begin{array}{ccc}
S & -\mathrm{i} D & 0 \\
\mathrm{i} D & S & 0 \\
0 & 0 & P
\end{array}\right),
\end{equation}
where the plasma parameters need to be summed for species $\alpha$ with charge $q_{\alpha}=\eta_\alpha\left|q_\alpha\right|$
\begin{equation}
\begin{aligned}
S & =\frac{1}{2}(R+L), \quad D=\frac{1}{2}(R-L) \\
R,L & =1-\sum_\alpha \frac{\omega_{\mathrm{p} \alpha}^2}{\omega^2} \frac{\omega}{\omega \pm \eta_\alpha \Omega_\alpha}, \quad P=1-\sum_\alpha \frac{\omega_{\mathrm{p} \alpha}^2}{\omega^2}.
\end{aligned}
\end{equation}
When the contributions from the ions are neglected, the final results become
\begin{equation}\label{eq:SDP}
S=1-\frac{X}{1-Y^2}, \quad D=-\eta\frac{X Y}{1-Y^2}, \quad P=1-X .
\end{equation}

If we choose the background magnetic field along 3-axes, and the angle between $\boldsymbol{k}$ and $\boldsymbol{B}$ is $\theta$, so that the wave equation can be written as
\begin{equation}\label{eq:wave_eq_cold}
\left[\begin{array}{ccc}
S-n^2 \cos ^2 \theta & -\mathrm{i} D & n^2 \cos \theta \sin \theta \\
\mathrm{i} D & S-n^2 & 0 \\
n^2 \cos \theta \sin \theta & 0 & P-n^2 \sin ^2 \theta
\end{array}\right]\left[\begin{array}{c}
E_x \\
E_y \\
E_z
\end{array}\right]=0.
\end{equation}
The dispersion relation for cold plasma is given by taking the coefficient determinant  to be zero $|\Lambda |=0$, i.e.
\begin{equation}\label{eq:dis_eq_cold}
A n^4-B n^2+C=0,
\end{equation}
where
\begin{equation}\label{eq:ABC_cold}
\begin{aligned}
& A=S \sin ^2 \theta+P \cos ^2\theta,  \\
& B=R L \sin ^2 \theta+P S\left(1+\cos ^2 \theta\right), \\
& C=P R L.
\end{aligned}
\end{equation}
By solving equation (\ref{eq:dis_eq_cold}), we obtain the dispersion relations for the cold plasma
\begin{equation}\label{eq:n_cold}
n_\pm^2=\frac{B \pm\left(B^2-4 A C\right)^{1 / 2}}{2 A},
\end{equation}
where `$\pm$' represents two natural modes.

\subsection{Polarization vector for cold plasma}\label{app:pol}
The wave polarization in the magnetized cold plasma can be described as the polarization vector. It is convenient to choose the wave vector in the 1–3 plane, and the basis vector is
\begin{equation}
\boldsymbol{\kappa}=(\sin \theta, 0, \cos \theta), \quad \boldsymbol{t}=(\cos \theta, 0,-\sin \theta), \quad \boldsymbol{a}=(0,1,0),
\end{equation}
where $\boldsymbol{\kappa}$ is the unit vector along the wave propagation direction, $\boldsymbol{t}$ is perpendicular to $\boldsymbol{\kappa}$ in the $\boldsymbol{k}-\boldsymbol{B}$ plane (1-3 plane) and $\boldsymbol{a}$ is perpendicular to the $\boldsymbol{k}-\boldsymbol{B}$ plane. For a given mode, the polarization vector is
\begin{equation}
\boldsymbol{e}_{\pm}=\frac{\left(L_{ \pm} \boldsymbol{\kappa}+T_{ \pm} \boldsymbol{t}+\mathrm{i} \boldsymbol{a}\right)}{\left(L_{ \pm}^2+T_{ \pm}^2+1\right)^{1 / 2}},
\end{equation}
where $T_{\pm}$ and $L_{ \pm}$ are transverse and longitudinal components, respectively. The wave polarization of the transverse part should be elliptical in general with $|T_{ \pm}|$ being the axial ratio. The handedness of the polarization ellipse depends on the sign of $T_{\pm}$: $T_{ \pm}>0$ for the right-hand polarization and $T_{ \pm}<0$ for the left-hand polarization.

The polarization vector can be constructed by cofactors of any column in the matrix $\Lambda$. For example, we can select the middle column. At the same time, for the convenience of comparison with equation (A8), we divide the entire equation by $\mathrm{i}$, that is
\begin{equation}
\lambda=\left(\begin{array}{c}
D\left(P-n^2 \sin ^2 \theta\right) \\
\mathrm{i}(A n^2-P S) \\
- D n^2 \sin \theta \cos \theta
\end{array}\right).
\end{equation}
To make these two vectors parallel to each other $\lambda=s\boldsymbol{e}_{\pm}$, we can get the longitudinal and transverse components
\begin{equation}
\begin{aligned}
&L=\frac{\left(P-n^2\right) D \sin \theta}{A n^2-P S},\\
&T=\frac{D P \cos \theta}{A n^2-P S}.
\end{aligned}
\end{equation}

\section{Wave dispersion in intrinsically relativistic plasma}\label{sec:appendixB}
\subsection{Dielectric tensor and dispersion relation for magnetospheric plasma}
The dispersion relation for relativistic pair plasma can be obtained in the same way as the case of cold plasma, but the covariant EM waves theory should be adopted. For a non-gyrotropic magnetar plasma, the dielectric tensor is \citep{Melrose1999}
\begin{equation}
\begin{aligned}
& K_{11}=K_{22}=1+\frac{1}{\beta_{\mathrm{A}}^2}\left(1+\frac{\Delta \beta^2}{z^2}\right), \\ 
& K_{33}=1-\frac{\omega_{\mathrm{p}}^2}{\omega^2} z^2 W(z)+\frac{\Delta \beta^2 \tan ^2 \theta}{\beta_{\mathrm{A}}^2 z^2} \\
& K_{12}=-K_{21}=-\mathrm{i} \eta \frac{\omega_{\mathrm{p}}^2}{\omega\Omega_{\mathrm{e}}}\\
& K_{13}=K_{31}=-\frac{\Delta \beta^2 \tan \theta}{\beta_{\mathrm{A}}^2 z^2}, 
\end{aligned}
\end{equation}
where the Alfvén speed $\beta_{\mathrm{A}}$ is
\begin{equation}
\beta_{\mathrm{A}}^2=\frac{\Omega_{\mathrm{e}}^2}{\omega_{\mathrm{p}}^2\langle\gamma\rangle}, 
\end{equation}
and the mean square speed $\Delta \beta$ is
\begin{equation}
    \Delta \beta^2=\frac{\left\langle\gamma \beta^2\right\rangle}{\langle\gamma\rangle}=1-\frac{\left\langle\gamma^{-1}\right\rangle}{\langle\gamma\rangle}.
\end{equation}

The average of any function $Q$ is written as
\begin{equation}
    \langle Q\rangle=\frac{1}{n}\int \mathrm{d} u Q g(u),
\end{equation}
where $g(u)$ is the particle distribution function (equation (\ref{eq:gu})). The relativistic plasma dispersion function (RPDF) is
\citep{MG99}
\begin{equation}
W(z)=\frac{1}{n} \int_{-\infty}^{\infty} \mathrm{d} u \frac{\mathrm{d} g(u) / \mathrm{d} u}{\beta-z}=\frac{1}{n} \int_{-1}^1 \mathrm{~d} \beta \frac{\mathrm{d} g(u) / \mathrm{d} \beta}{\beta-z},
\end{equation}
with $z=\omega /k c \cos \theta$. In the non-relativistic limit, $\langle\gamma\rangle\rightarrow 1$ and $z^2 W(z)\rightarrow 1$, the dispersion relation is the same as that of cold plasma.

As with cold plasma, the wave equation can be obtained based on the dielectric tensor\footnote{The 22 component was mistakenly written as $\Lambda_{22}=\Lambda_{11}+\sin^2 \theta/z^2$ in \cite{Melrose1999,MG99}, and it has been corrected in \cite{Rafat2019}.} \citep{Melrose1999,MG99,Rafat2019}. The non-zero components of $\tilde{\Lambda}$ are \citep{Rafat2019}
\begin{equation}
\begin{aligned}
& \tilde{\Lambda}_{11}=\tilde{S}-\tilde{n}^2 \cos ^2 \theta, \\ 
& \tilde{\Lambda}_{12}=-\tilde{\Lambda}_{21}=\mathrm{i} \tilde{D}, \\
& \tilde{\Lambda}_{22}=\tilde{\Lambda}_{11}-n^2\sin ^2 \theta\\
& \tilde{\Lambda}_{33}=\tilde{P}-\tilde{n}^2 \sin ^2 \theta, \\ 
& \tilde{\Lambda}_{13}=\tilde{\Lambda}_{31}=\tilde{n}^2 \cos \theta \sin \theta,
\end{aligned}
\end{equation}
where $\tilde{n}^2=bn^2$ and $b=1-\Delta \beta^2/\beta_{\mathrm{A}}^2$. The modified plasma parameters are \citep{Melrose2004}
\begin{equation}\label{eq:SDP_rel}
\begin{aligned}
    \tilde{S}=1+\frac{\tilde{X}}{\tilde{Y}^2}, \quad \tilde{D}&=\eta \frac{\tilde{X}}{\tilde{Y}}, \quad \tilde{P}=1-\frac{\omega_{\mathrm{p}}^2}{\omega^2} \sec ^2 \theta W(\sec \theta),\\
    \tilde{X}&=\frac{\omega_{\mathrm{p}}^2}{\langle\gamma\rangle \omega^2}, \quad \tilde{Y}=\frac{\Omega_{\mathrm{e}}}{\langle\gamma\rangle \omega}.
\end{aligned}
\end{equation}
The weak anisotropy limit $z=\sec \theta$ is used in $\tilde{P}$. The similar expressions are given in \cite{Melrose2004}, except the component $\tilde{\Lambda}_{22}$. The dispersion relation for relativistic plasma can be obtained from $|\tilde{\Lambda}|=0$, i.e.
\begin{equation}\label{eq:dis_eq_rel}
\tilde{A} n^4-\tilde{B} n^2+\tilde{C}=0,
\end{equation}
where
\begin{equation}\label{eq:ABC_rel}
\begin{aligned}
& \tilde{A}=b(b\cos ^2\theta+\sin ^2 \theta)(\tilde{S} \sin ^2 \theta+\tilde{P} \cos ^2\theta), \\
& \tilde{B}=b(\tilde{S}^2-\tilde{D}^2) \sin ^2 \theta+\tilde{P} \tilde{S}\left(2b\cos ^2+\sin ^2 \theta\right), \\
& \tilde{C}=\tilde{P} (\tilde{S}^2-\tilde{D}^2).
\end{aligned}
\end{equation}

\subsection{Polarization vector for magnetospheric plasma}

The polarization vector can be constructed similarly in Appendix A.2. Here we only list the results of transverse and longitudinal components
\begin{equation}
\begin{aligned}
&L=\frac{\left(P-\tilde{n}^2\right) \tilde{D} \sin \theta}{(\tilde{S} \sin ^2 \theta+\tilde{P} \cos ^2\theta) \tilde{n}^2-\tilde{P} \tilde{S}},\\
&T=\frac{\tilde{D} \tilde{P} \cos \theta}{(\tilde{S} \sin ^2 \theta+\tilde{P} \cos ^2\theta) \tilde{n}^2-\tilde{P} \tilde{S}}.
\end{aligned}
\end{equation}
From equation (\ref{eq:ABC_rel}), the transverse component satisfies a quadratic equation 
\begin{equation}\label{eq:T_rel}
T^2-\mathcal{R} T-d=0,
\end{equation}
where
\begin{equation}\label{eq:R_rel}
\begin{aligned}
    &\mathcal{R}=\frac{\left(\tilde{P} \tilde{S}+b(\tilde{D}^2-\tilde{S}^2)\right) \sin ^2 \theta}{b\tilde{P} \tilde{D} \cos \theta},\\
    & d = \frac{1+b+(b-1)\cos 2\theta}{2b}.
\end{aligned}
\end{equation}
The solutions are
\begin{equation}
T=T_{ \pm}=\frac{1}{2}\left[\mathcal{R}\pm\left(\mathcal{R}^2+4d\right)^{1 / 2}\right]. 
\end{equation}
For the typical magnetospheric plasma ($\Omega_{\mathrm{e}}\gg\omega \gg \omega_{\mathrm{p}}$ and $\langle\gamma\rangle\gg1$), $b\sim 1$ is a good approximation. Thus, equation (\ref{eq:T_rel}) and the parameter $\mathcal{R}$ (equation \ref{eq:R_rel}) have the same form as for cold plasma, where the plasma parameters need to be corrected for relativistic distribution (equation \ref{eq:SDP_rel}). 
\end{appendix}

\end{document}